%% file: main.tex
\documentclass[runningheads,]{llncs}
\usepackage{graphicx}
\usepackage{amsmath}
\usepackage{amsfonts}
\usepackage{comment}
\usepackage{mathtools}
\usepackage{multirow}
\usepackage{booktabs,rotating,tabularx}
\usepackage{subfigure}

\DeclareMathOperator{\Tr}{Tr}

\newcommand\st[1]{}

\begin{document}
%
\newcommand{\psge}{{\sc{PSGE}}}
\newcommand{\mat}{\mathrm}
\newcommand{\pdv}[2]{\frac{\partial {#1}} {\partial {#2}}}

\title{Pure Spectral Graph Embeddings: Reinterpreting Graph Convolution \\for Top-N Recommendation}

\titlerunning{Reinterpreting Graph Convolution for Top-N Recommendation}
%
\author{Edoardo D'Amico\orcidID{0000-0002-8262-7207}\inst{1,2} \and \\Aonghus Lawlor\orcidID{0000-0002-6160-4639}\inst{1,2} \and Neil Hurley\orcidID{0000-0001-8428-2866}\inst{1,2}}
\authorrunning{E. D'Amico et al.}
%
\institute{Insight Centre for Data Analytics, Dublin, Ireland\\
\email{\{name.surname\}@insight-centre.org}
\and University College of Dublin, Ireland}
\maketitle              
%

\input{00-Abstract.tex}
\input{01-Introduction}
\input{02-Preliminaries}
\input{03-Methodology}

\input{04-Experiments}
\input{05-Conclusion}
\subsubsection*{Acknowledgments}
This research was supported by Science Foundation Ireland (SFI) under Grant Number SFI/12/RC/2289\_P2.
\bibliographystyle{splncs04}
\bibliography{bibliography2}

%
%

\end{document}

%% file: 00-Abstract.tex
\begin{abstract}
The use of graph convolution in the development of recommender system algorithms has recently achieved state-of-the-art results in the collaborative filtering task (CF).
While it has been demonstrated that the graph convolution operation is connected to a filtering operation on the graph spectral domain, the theoretical rationale for why this leads to higher performance on the collaborative filtering problem remains unknown.
The presented work makes two contributions. First, we investigate the effect of using graph convolution throughout the user and item representation learning processes, demonstrating how the latent features learned are pushed from the filtering operation into the subspace spanned by the eigenvectors associated with the highest eigenvalues of the normalised adjacency matrix, and how vectors lying on this subspace are the optimal solutions for an objective function related to the sum of the prediction function over the training data. 
Then, we present an approach that directly leverages the eigenvectors to emulate the solution obtained through graph convolution, eliminating the requirement for a time-consuming gradient descent training procedure while also delivering higher performance on three real-world datasets. 
\keywords{Collaborative filtering  \and Graph convolution \and Spectral methods.}
\end{abstract}

%% file: 01-Introduction.tex
\section{Introduction}
Graph convolutional networks (GCN) are a form of deep learning network which leverages the structural information in a graph representation of the training data \cite{welling2016semi}. 
The convolutional layers of the network aggregate each nodal feature with those of its neighbours in the graph. By constructing a network of $h$ convolutional layers, the node embedding becomes dependent on the features of nodes that are $h$-hops away from it in the network. 
\st{GCNs have been applied to recommendation, for instance in the NGCF~\cite{wang2019neural} model, that employs a fully convolutional network similar to those used in other machine learning applications, to propagate user and item embeddings through the user-item interaction graph. Training of the embeddings is driven by a Bayesian Personalised Ranking~\cite{rendle2012bpr} loss function at the output of the network.}
We focus on the  LightGCN algorithm~\cite{he2020lightgcn} that has received a lot of attention recently due to the fact that it has demonstrated that, given only user-item interaction data without rich user and item features, the convolutional layers can be greatly simplified. In particular, it argues that the non-linear activation  and the trainable weights of the full GCN can be removed from the convolution without any degradation to the accuracy of the model and a substantial saving in the complexity of training. LightGCN has been shown to obtain state-of-the-art performance in terms of top-$N$ performance measures on a number of recommender datasets. In this paper, we address the question of why LightGCN achieves good performance, despite its simple convolutional layers. While it is surely true that LightGCN is less complex than a regular GCN \cite{welling2016semi}, it is also true that it requires training by gradient descent and that each update to the model parameters is much more complex than the updates of standard matrix factorisation algorithms, such as BPR. Hence, we ask  if LightGCN is fundamentally better at capturing features in the dataset that a standard matrix factorisation model will miss. 
We show that, without the non-linear activation functions, the convolutions of LightGCN can be understood as graph filters that have the effect of generating features that are largely embedded in a subspace spanned by the eigenvalues of the normalised interaction matrix corresponding to its largest eigenvalues. We show why this is a suitable subspace in which to find quality solutions to solve the top-$N$ recommendation problem. With this spectral interpretation of LightGCN, we proceed to build spectral recommender model, which we call \textit{Pure Spectral Graph Embeddings} (PSGE) that leverages the principles behind LightGCN, while having a closed-form solution that can be found through an eigen-decomposition of the interaction matrix, rather than through a gradient descent algorithm. Given that fast algorithms for eigen-decomposition of sparse matrices are available \cite{lanczos1996}, \psge{} can be learned in a fraction of the time that it takes to train LightGCN. We demonstrate that PSGE out-performs LightGCN on a number of recommendation datasets. We also test its performance against the other leading linear algorithms in the literature and show that PSGE can be configured to achieve high recommendation performance while reducing the popularity bias that is evident in these other similar algorithms.

\st{The remainder of the paper is structured as follows: In Section 2, we establish the problem setting and notation. Following this, in Section 3, we discuss how LightGCN acts as a high-pass filter on the user and item embeddings and, in Section 4, we describe the optimisation objective that such a filtering targets. In Section 5, we use the learnings from LightGCN to develop a novel spectral algorithm, which is evaluated in Section 6 and shown to out-perform a number of other comparable algorithms from the state-of-the-art.}

%% file: 02-Preliminaries.tex
\section{Preliminaries}
Let $\mathcal{U}$ be a set of users of size $|\mathcal{U}|=U$ and $\mathcal{I}$ be a set of items of size $|\mathcal{I}|=I$.
Given a $U \times I$  interaction dataset $\mat{R}=\{r_{ui}\}$ where $r_{ui}$ represents implicit feedback given by user $u$ on item $i$,  the top-$N$ recommendation problem is to recommend a set of $N>0$ items that the system predicts are relevant to a given user $u$. Typically, a prediction function computes a relevance score $\hat r_{ui}$, the items are ranked according to $\hat r_{ui}$ and the top items in this ordering are recommended.   We focus on latent space methods, where, for each user and item, a $f$-dimensional embedding, denoted respectively as $\mathbf{p}_u$ and $\mathbf{q}_i$ is learned from the interaction data, and the prediction function is the inner product of the user and item embeddings.
Write $\mat{P}$ for the $U \times f$ matrix whose rows are the user embeddings $\mathbf{p}_u$ and $\mat{Q}$ for the $I \times f$ matrix whose rows are the item embeddings $\mathbf{q}_i$. 

Graph convolution methods interpret user-item interactions as edges of a graph. More formally, the interaction data $\mat{R}$ can be represented as an undirected bipartite graph $\mathcal{G}_R = (\mathcal{V},\mathcal{E})$ with nodes $\mathcal{V} = \mathcal{U} \cup \mathcal{I}$ and edges $\mathcal{E}=\{(u,i) | u \in \mathcal{U}, i \in \mathcal{I}, r_{ui} \ne 0\}$ connect user nodes $u$ to item nodes $i$ whenever there is an interaction between them in $\mat{R}$. 
\subsection{Graph Signals and the Graph Fourier Transform}
Given an undirected weighted graph of order $n$, with adjacency matrix $\mat{A} = \{a_{ij}\} \in \mathbb{R}$, a  signal over the graph is a function $f: \mathcal{V }\rightarrow \mathbb{R}$. For any signal, we can form the $n$-dimensional vector $\mathbf{x} \in \mathbb{R}^n$ such that the $i^{th}$ component of $\mathbf{x}$ represents the value of the signal at the $i^{th}$ vertex of $\mathcal{V}$. 
Notice how concatenating the embedding matrices, $\mat{P}$ and $\mat{Q}$ into a single $(U+I)\times f$ dimensional matrix $\mat{X}$, 
we obtain a matrix in which each \emph{column} represents a signal over the bipartite graph $\mathcal{G}_R$. From this alternative point of view, the learning process involves the learning of $f$ different signals over the graph which can be thought as \textit{latent features} constructing the user and item representations.

A graph convolution operation on a signal is a weighted sum of the signal at a node with its values in a neighbourhood of up to $(n-1)$-hops from the node  and can be represented as a polynomial over a propagation matrix $\mat{S}$ with weights $g_i$: 
\[
	conv(\mathbf{x},\mathbf{g}) = \sum_{i=0}^{n-1} g_i \mat{S}^i \mathbf{x} \equiv \mathbf{g} * \mathbf{x} \,.
\]
A common choice for the propagation matrix $\mat{S}$ is the normalised Laplacian $\mat{\Delta}$, of a graph with adjacency $\mat{A}$, defined as $\mat{\Delta}=\mat{I}-\mat{D}^{-1/2}\mat{A}\mat{D}^{-1/2}\,,$
where $\mat{D}$ is the diagonal matrix of node degrees with diagonal elements $d_{ii} = \sum_{j} a_{ij}$. 
$\mat{\Delta}$ is a symmetric positive semi-definite matrix, so that its eigenvalues $\lambda_i$ are non-negative and its eigenvectors form an orthogonal basis, allowing the decomposition,
$
\mat{\Delta} = \mat{U} \mat{\Lambda} \mat{U}^{\top},
$
where $\mat{U}$ is the matrix whose columns are the $n$ orthonormal eigenvectors and $\mat{\Lambda} = \mathrm{diag}(\lambda_1, \ldots, \lambda_n)$ is the diagonal matrix of the eigenvalues, assumed ordered such that $0=\lambda_1 \le \lambda_2 \le \dots \le \lambda_n$. 
Note that, while  $\mat{\Delta}$ is a common choice, any real symmetric matrix associated with the graph can be chosen to define the graph spectrum.

The \textit{Graph Fourier Transform}~\cite{shuman2013emerging} $\hat{\mathbf{x}}$ of a graph signal $\mathbf{x}$ is defined as its projection into the eigenvector basis $\mat{U}$, i.e., $\hat{\mathbf{x}}= \mat{U}^{\top}\mathbf{x}$, with inverse operation defined as $\mathbf{x} = \mat{U}\mathbf{\hat{x}}$. In the Fourier domain of the eigenvector basis, a convolution is a simple element-wise multiplication, such that $\widehat{\mathbf{g} * \mathbf{x}} = \hat{\mathbf{g}}   \hat{\mathbf{x}}$, where\footnote{A polynomial $p(\mat{A})$  has the same eigenvectors as $\mat{A}$, with eigenvalues given by $p(\lambda)$, where $\lambda$ is an eigenvalue of $\mat{A}$.},
\[
	\hat g_i = \hat g_i(\lambda_i) = \sum_{j=0}^{n-1} g_j \mat{U}_j^i
\]
This shows that the spectral coefficient $\hat{x}_i(\lambda_i)$ reflecting the correlation of the signal $\textbf{x}$ with the $i^{\rm th}$ eigenvector, is scaled by $\hat g_i(\lambda_i)$ and hence $\hat{\mathbf{g}}$ can be thought of as a spectral filter, which can enhance or diminish certain frequencies of the signal. We can understand the impact of the convolution on a signal most easily by studying 
$\hat{g}_i(\lambda_i)$ in the Fourier domain.

%
%

%

In the collaborative filtering problem only the historical user-item interactions are available, making impossible to seed the initial representations (signals) for the user and item nodes. In \cite{wang2019neural}, it is proposed to initialise the node embeddings as free parameters and learn jointly the representations and filters from the training data. The complexity of this approach has been shown to downgrade the quality of the user and item representations learnt~\cite{chen2020revisiting,he2020lightgcn}. To overcome this problem, in~\cite{he2020lightgcn}, the ``light convolution'' method, LightGCN, is proposed, where the only free parameters correspond to the user and item representations. In the following, we show that the chosen propagation matrix  corresponds to a fixed high-pass filter in the spectral domain defined by the normalised adjacency matrix. 

\section{LightGCN as a High-Pass Filter}
\label{sec:lightgcn}
LightGCN \cite{he2020lightgcn} is a state-of-the-art graph convolution model for the top-$N$ recommendation task. It uses $\mat{S}=\mat{D}^{-1/2}\mat{A}\mat{D}^{-1/2}$ as a propagation matrix to exchange information along the edges of the graph, where $\mat{A}$ is the adjacency matrix of the user-item interaction graph $\mathcal{G}_R$.  At the first step, the latent features of user and items (signals) $\mat{X}^{(0)} = [\mat{P}^{(0)};\mat{Q}^{(0)}] \in \mathbb{R}^{(U+I)\times f}$ are randomly initialised and then updated at every convolution step as $\mat{X}^{(k)} = \mat{S}\mat{X}^{(k-1)}\,.$
The final user and item latent features are then computed as a weighted combination of the signals at each convolution step:
\begin{align*}
\begin{aligned}
    \mat{X} &= \alpha_0\mat{X}^{(0)} + \ldots + \alpha_k\mat{X}^{(k)} 
    = \big(\alpha_0\mat{I} + \alpha_1\mat{S}+ \ldots + \alpha_k\mat{S}^{k}\big)\mat{X}^{(0)} 
\end{aligned}
\end{align*}
The authors reported that learning the signals and the coefficients $\alpha_i$ jointly lead to worse results than assigning the uniform weights, $\alpha_0 = \alpha_1 = \ldots = \alpha_k = 1/(k+1)$.

By carrying out an analysis over the spectrum defined from the symmetric normalised Laplacian $\mat{\Delta}$, in~\cite{shen2021how}, the convolution operation  is shown to correspond to a low-pass filter. If we instead consider the spectrum defined by the propagation matrix $\mat{S} = \mat{I} - \mat{\Delta}$ -i.e. the symmetric normalised adjacency matrix- we show how the convolution correspond to an high-pass filter.
\begin{align*}
\begin{split}
\hat{g_i}(\lambda_i) & = \frac{1}{k+1}(1+\lambda_i+\lambda_i^2+\ldots+\lambda_i^k)\\
\end{split}
\end{align*}
where $\lambda_i$ are the eigenvalues of $\mat{S}$. Applying the sum of  the geometric series, we can express it more concisely as:
\begin{equation}
    \hat{g}(\lambda) = 
    \begin{cases}
    \frac{1}{k+1}\frac{1-\lambda^{k+1}}{1 - \lambda} & \lambda<1\\
    1 & \lambda=1\\
    \end{cases}
\end{equation}
\input{tex_figures/LightGCN_filters}
In Fig. \ref{fig:LightGCN_filters}, $\hat{g}_i$ is plotted against $\lambda_i$ for different values of the convolution depth $k$. It illustrates that the convolution acts as a high-pass filter over the spectrum defined by S, reducing the strength of the lower frequencies, with stronger filtering as $k$ increases. This implies that the convolution operation transforms the input signals so that they are focused in a subspace spanned by the eigenvectors corresponding to the high eigenvalues of the normalised adjacency ${\mat{S}}$. 

%% file: tex_figures/LightGCN_filters.tex
\begin{figure}
    \begin{center}
    \includegraphics[width=0.6\columnwidth]{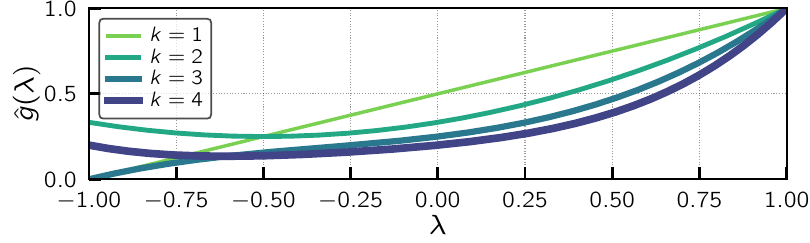}
    \caption{LightGCN spectral filter for different values of $k$.}
    \label{fig:LightGCN_filters}
    \end{center}
\end{figure}

%% file: 03-Methodology.tex
\section{A Spectral Interpretation of LightGCN}
\label{sec:spectral_interpretation}
We have highlighted how the final latent features learnt from LightGCN are substantially contained in the span of the largest eigenvectors of the normalised adjacency matrix. The theoretical underpinnings for why this is beneficial for the recommendation task are discussed in this section. We show that leveraging vectors lying in this subspace as latent features, leads to the optimisation of a target function which is a weighted summation of the prediction function over the training data.


When the problem is rating prediction, the goal of a recommendation algorithm is to learn a prediction matrix, 
$\mat{\hat{R}}$ which well approximates the rating matrix $\mat{R}$. This can be formulated in terms of finding $\mat{\hat{R}}$ which is close to $\mat{R}$ in the Frobenius norm:
\begin{align*}
	&\min_{\hat{R}} \big(\| \mat{R} - \mat{\hat{R}}\|_F^2 \big) = \min_{\hat{R}}\big(\Tr((\mat{R} - \mat{\hat{R}})(\mat{R} - \mat{\hat{R}})^T)\big)\\
	&= \min_{\hat{R}}\big(\Tr(\mat{R}\mat{R}^T) - 2\Tr(\mat{R}\mat{\hat{R}}^T) + \Tr(\mat{\hat{R}}\mat{\hat{R}^T})\big)
	= \min_{\hat{R}}\big( - 2\Tr(\mat{R}\mat{\hat{R}}^T) + \|\mat{\hat{R}}\|^2_F\big)\,.
\end{align*}
However, when the problem is top-$N$ recommendation where the requirement is to learn a score to sort the items in order of preference, the scale  of the prediction function is irrelevant to the order and can be fixed to any arbitrary value. Hence, we can write the target objective as:
\begin{align*}
	&\max_{\mat{\hat{R}} }\big(\Tr(\mat{R}\mat{\hat{R}}^T) \big) \quad\mbox{ s.t. } \| \mat{\hat{R}} \|_F^2 \mbox{ is fixed}.
\end{align*}
In fact, for implicit binary datasets, the trace has a natural interpretation as the sum of the predictions over the positive interaction data. Furthermore, note that the trace can be written as a quadratic form over the adjacency matrix of the user-item interaction graph.
\begin{proposition}
The quadratic form $\mathcal{Q}_A(\mathbf{x})$ induced by the adjacency matrix of the user-item interaction graph, on a signal $\mathbf{x} = [\mathbf{p};\mathbf{q}]$, corresponds to twice the sum of the prediction function over all positive interactions in the training dataset.
\begin{align*}
    	\mathcal{Q}_A(\mathbf{x}) &= \mathbf{x}^T \mat{A} \mathbf{x} = \sum_{\ell,k} a_{\ell k} x_\ell x_k
 = 2\sum_{\{(u,i) | r_{ui} \ne 0\} }  p_u q_i \,.
\end{align*}
This can be generalised to a $(U+I)\times f$ matrix $\mat{X}$ of  $f$ signals as:
\begin{align}
\label{eq:psge_tf}
  \mathcal{Q}_A(\mat{X}) &= \nonumber \Tr\big(\mat{X}^T \mat{A} \mat{X}\big) = \Tr\big( \mat{A} \mat{X}\mat{X}^T\big)
  =2 \Tr\big(\mat{P}^T \mat{R} \mat{Q}\big) = 2 \sum_{\{(u,i) | r_{ui} \ne 0\}}  \mathbf{p}^T_u \mathbf{q}_i \,.
\end{align}
such that the sum of the prediction functions over the training data positive interactions is the trace of a quadratic form on the interaction data. 
\end{proposition}
This is an intuitive objective for the top-$N$ recommendation task,as opposed to the rating prediction task.
Note that  any rank $f$, symmetric matrix $\mat{X}\mat{X}^T$ can be written as $\mat{Y}\mat{\Sigma}\mat{Y}^T$ where $\mat{Y}$ is orthogonal (i.e. $\mat{Y}^T\mat{Y} = \mat{I}_f$) and $\mat{\Sigma}$ is a $f \times f$ diagonal matrix. So we can equivalently write the trace as  $\mathcal{Q}_A(\mat{X}, \mat{\Sigma}) = \Tr\big(\mat{X}^T \mat{A} \mat{X}\mat{\Sigma}\big)$, for orthogonal $\mat{X}$.
As such, we recognise that the problem of learning $f$ signals (latent features) to construct the user and item embeddings which maximises the sum of the prediction function over the training data is solved by the generalised Rayleigh-Ritz theorem \cite{magnus_1998}.



\begin{theorem}[Rayleigh-Ritz]
\label{th:Rayleigh-Ritz}
For a real symmetric $n \times n$ matrix $\mat{A}$:
\begin{align*} 
    &\max_{\mat{X}}\{ \Tr\big(\mat{X}^T \mat{A} \mat{X}\big) \: \mathrm{s.t.}\: \mat{X}^T\mat{X} = \mat{I}_f \}
    = \lambda_{1}+\dots+\lambda_{f} 
\end{align*}
and the maximising matrix is $\mat{X}=[\mathbf{v}_1, \dots, \mathbf{v}_{f}]$ where $\lambda_i$  are the $f$ largest eigenvalues of $\mat{A}$ and $\mathbf{v}_i$ the corresponding orthonormal eigenvectors. 
Furthermore \cite{AbsilMahonySepulchre2009}, the quadratic form $\Tr\big(\mat{X}^T \mat{A} \mat{X}\mat{\Sigma}\big)$, where $\mat{\Sigma} = \mathrm{diag}(\sigma_i)$ is a fixed diagonal matrix, is optimised by the same matrix of orthonormal eigenvectors, such that
\begin{align} 
     &\Tr\big(\mat{X}^T \mat{A} \mat{X}\mat{\Sigma} \big)  = \sum_{i=1}^f \lambda_i \sigma_i\,.
 \end{align}
\end{theorem}


\subsection{Inverse Propensity Control}
It is well recognised that recommender system datasets tend to exhibit biases in the manner in which the interaction data is observed \cite{chen2020bias}. Propensity scoring provides one means of taking such biases into account during model learning \cite{schnabel2016recommendations,zhu2020unbiased}. The propensity score is an estimate of the probability that any particular interaction is observed. The contribution of each observed interaction to the loss function is multiplied by its inverse propensity score, prior to model learning.   The symmetric normalised adjacency matrix of the LightGCN method can be viewed as an inverse propensity weighted adjacency. In particular, each observed $a_{ui}$ is weighted by a term depending on
the user and item degrees: $d_u^{-1/2} d_i^{-1/2}$.  The quadratic form over this normalised matrix, ${\mat{\tilde{A}}}$ is a weighted sum of the predictions on the training data, where each prediction is down-weighted according to the user and item degree:
\begin{align*}
    \mathcal{Q}_{\tilde{A}}(\mathbf{x}) =
     \Tr\big(\mat{X}^T \tilde{\mat{A}} \mat{X}\big) =
    \Tr\big(\mat{X}^T \mat{D}^{-1/2}AD^{-1/2} \mat{X}\big)
    = 2 \sum_{\{(u,i) | r_{ui} \ne 0\}}  \frac{1}{d_u^{1/2} d_i^{1/2}} \mathbf{p}^T_u \mathbf{q}_i  \,.
\end{align*}
Without such normalisation, the target function can be trivially maximised by giving larger embedding weights to the users and items with many interactions in the training set. 
The normalisation should therefore have the effect of increasing the embedding weights of unpopular items and users with short profiles. 

The eigenvectors with largest eigenvalues of the normalised adjacency provide the optimal solution for this modified target objective. Hence, we can conclude that to a large extent the LightGCN method is effective because the convolution focuses on embeddings that are largely contained in the subspace spanned by the eigenvectors of largest eigenvalues of the normalised adjacency; and that these eigenvectors provide an optimal solution to the target objective of maximising the propensity-weighted sum of the predictions over the training data.
It is noteworthy that, in the maximisation of the quadratic form, each eigenvector contributes proportionally to its associated eigenvalue (Theorem \ref{th:Rayleigh-Ritz}), meaning that is reasonable to assume that the latent features associated to higher eigenvectors should have more weight with respect to those  associated to lower eigenvectors. The shape of the high pass filter employed by LightGCN throughout the learning process, Fig. \ref{fig:LightGCN_filters}, can deliver such a spectrum.

\section{Pure Spectral Graph Embeddings Model}
\label{sec:psge_models}
Given the interpretation of LightGCN in terms of the spectrum of the adjacency matrix, it is worth asking if spectral methods can be developed that are competitive with LightGCN on accuracy. 
Firstly, we show how the PureSVD \cite{cremonesi2010performance} can be interpreted under a trace maximisation problem. Explaining the doubts presented in the original paper regarding how a method devised for rating prediction is performing so well with implicit feedbacks.


\subsection{PureSVD}
\label{subsec:PureSVD}
$\mathcal{Q}_A(\mat{X}, \mat{\Sigma})$ is maximised when  $\mat{X} = [\mat{P};\mat{Q}]$ are the eigenvectors of $\mat{A}$ and the prediction function is then $\mat{\hat{R}} = \mat{P}\mat{\Sigma}\mat{Q}^T$, where $\mat{\Sigma} = \mathrm{diag}(\sigma_i)$.
Writing $\mathbf{u}$ for a $U \times 1$-dimensional eigenvector of $\mat{R}\mat{R}^T$, and  $\mathbf{v}$ for the corresponding $I \times 1$-dimensional eigenvector of $\mat{R}^T\mat{R}$, with eigenvalue $\lambda^2\ge0$, the eigenvectors of $\mat{A}$ are $\mathbf{x} = \frac 1 {\sqrt{2}}[\mathbf{u};\mathbf{v}]$ and $\mathbf{x} = \frac 1 {\sqrt{2}}[\mathbf{u};-\mathbf{v}]$ with eigenvalues $\pm\lambda$.  Moreover, 
$
	\mathbf{u} = \mat{R}\mathbf{v} / \lambda 
$
or, gathering all $f$ eigenvectors into the columns of  $\mat{P}$ and $\mat{Q}$, we have $\mat{P} = \mat{R} \mat{Q} \mat{\Lambda}^{-1}$ where $\mat{\Lambda} = \mathrm{diag}(\lambda_i)$.
The eigenvectors $\mathbf{u}$ and $\mathbf{v}$ can be obtained from a singular value decomposition of $\mat{R}$ \cite{kunegis2015exploiting}. 

The coefficients $\sigma_i$ in the above expression can be interpreted as a  weight given to each of the $f$ signals from which the embedding is formed. Now, $\| \hat{R} \|^2  = \| \mat{P}\mat{\Sigma}\mat{Q}^T \|^2 = \sum_{i=1}^f \sigma_i^2$ and it follows that the best choice of $\sigma_i$ to maximise $\sum_{i=1}^f \lambda_{i} \sigma_i$, under a fixed constraint on its norm is $\sigma_i \propto  \lambda_i$. The prediction function is then $\mat{P}\mat{Q}^T \mat{\Sigma} = \mat{R}\mat{Q}\mat{\Lambda}^{-1}\mat{\Lambda}\mat{Q}^T = \mat{R}\mat{Q}\mat{Q}^T$
which is exactly the PureSVD method. 
We have arrived at this method through trace maximisation under a norm constraint on the prediction function, as opposed to the Frobenius norm minimisation approach, appropriate for rating prediction. The trace maximisation perspective allows for the development of other methods, which differ from PureSVD in the manner in which the norm of the prediction matrix is constrained.

\subsection{Propensity Weighted Norm Constraint}
Given that we wish to control the size of the embeddings of highly active users or popular items, it is useful to consider a constraint on the prediction function that controls the embedding size in proportion to the degree in the interaction dataset.  In particular, we consider the following trace maximisation problem:
\begin{align*}
	&\max_{\mat{\hat{R}} }\big(\Tr(\mat{R}\mat{\hat{R}}^T) \big) \quad\mbox{ s.t. } \| \mat{D}_U^\alpha \mat{\hat{R}}  \mat{D}_I^\beta \|^2_F\quad\text{is fixed}\,.
\end{align*}
where $\mat{D}_U$ is the diagonal matrix of user degrees and $\mat{D}_I$ the diagonal matrix of item degrees.
By scaling the contribution of each prediction $\hat{r}_{ui} = \mathbf{p}_u^T \mathbf{q}_i$ in this fixed norm by the degrees $d_u^\alpha d_i^\beta$,  the effect, as $\alpha$ and $\beta$ get larger will be that the size of high degree embeddings gets smaller. 
Here we have generalised the exponent used in LightGCN to allow for two tuneable parameters $\alpha$ and $\beta$ such that we can explicitly control  the propensity score attributed to the users and items. 
With a change of variables $\mat{\tilde{P}} = \mat{D}_U^\alpha \mat{P}$ and $\mat{\tilde{Q}} = \mat{D}_I^\beta \mat{Q}$, we have

\begin{align*}
	\Tr(\mat{R} \mat{\hat{R}}^T) &= \Tr(\mat{R} \mat{Q}\mat{P}^T ) = \Tr(\mat{R} \mat{D}_I^{-\beta}\mat{\tilde{Q}}\mat{\tilde{P}}^T\mat{D}_U^{-\alpha})\\
	&=\Tr((\mat{D}_U^{-\alpha}\mat{R} \mat{D}_I^{-\beta})\mat{\tilde{Q}}\mat{\tilde{P}}^T) = \frac 1 2 \Tr(\mat{\tilde{X}}^T(\mat{D}\mat{A} \mat{D})\mat{\tilde{X}})
\end{align*}
where $\mat{D}$ is the $(U+I)\times(U+I)$ diagonal matrix $diag(\mat{D_U^{-\alpha}},\mat{D_I^{-\beta}})$.
Writing $\mat{\tilde{A}} = \mat{D}\mat{A} \mat{D}$ as the normalised adjacency, the trace is maximised by choosing $\mat{\tilde{P}}$ and $\mat{\tilde{Q}}$ from the PureSVD solution on the normalised interaction matrix $\mat{\tilde{R}} =  \mat{D}_U^{-\alpha}\mat{R} \mat{D}_I^{-\beta}$.

Having found $\mat{\tilde{P}}$ and $\mat{\tilde{Q}}$, one way to proceed  is to rescale them back to the required factors $\mat{P}$ and $\mat{Q}$, to obtain a fully popularity-controlled prediction matrix.  On the other hand,  it is well known that to achieve high recommendation accuracy, some popularity bias in the model's predictions is required~\cite{steck2011item}. 
So, instead, we complete the prediction function by noting that, since the embeddings are produced from the SVD of $\mat{\tilde{R}}$:
\\
\begin{enumerate}
    \item   \label{item1}       $\mat{\tilde{P}} = \mat{\tilde{R}}\mat{\tilde{Q}}\mat{\tilde{\Lambda}}^{-1}$, and
    \item  \label{item2}       $ \mat{\tilde{P}}\mat{\tilde{\Lambda}} \mat{\tilde{Q}}^T = \mat{\tilde{R}}\mat{\tilde{Q}}\mat{\tilde{Q}}^T \approx \mat{\tilde{R}}$.
\end{enumerate}   
Hence 
\begin{align*}           
\mat{D}_U^{-\alpha}\mat{R} \mat{D}_I^{-\beta}\mat{\tilde{Q}}\mat{\tilde{Q}}^T &\approx \mat{D}_U^{-\alpha}\mat{R} \mat{D}_I^{-\beta} &\text{(from (\ref{item1}) and (\ref{item2}) above)}\\
\mat{R} \mat{D}_I^{-\beta}\mat{\tilde{Q}}\mat{\tilde{Q}}^T &\approx \mat{R} \mat{D}_I^{-\beta}&\text{(dividing by }\mat{D}_U^{-\alpha})\\
\mat{R} \mat{D}_I^{-\beta}\mat{\tilde{Q}}\mat{\tilde{Q}}^T\mat{D}_I^{\beta}\ &\approx \mat{R}&\text{(multiplying by }\mat{D}_I^{\beta}) \,.
\end{align*}
So, we set $\mat{\hat{R}} = \mat{R} \mat{D}_I^{-\beta}\mat{\tilde{Q}}\mat{\tilde{Q}}^T\mat{D}_I^{\beta}\,,$
as the prediction matrix that directly approximates the observed interaction data, while being constructed in a manner that accounts for user and item propensity.
It is worth noting that, although $\alpha$ does not appear explicitly in this formula, the eigenvectors in $\mat{\tilde{Q}}$ depend on $\alpha$, as they are computed from the user- and item-degree normalised matrix. In fact, using (\ref{item1}) we can equivalently write the prediction matrix as
\begin{align}
\label{eq:psge_prediction}
\mat{\hat{R}} = \mat{D}_U^{\alpha}\mat{\tilde{P}}\mat{\Lambda}\mat{\tilde{Q}}^T\mat{D}_I^{\beta}\,.
\end{align}
 We name this method \textit{Pure Spectral Graph Embeddings} (\psge{}).

%% file: 04-Experiments.tex
\section{Experiments}
\input{tex_tables/trainval_perf_table_2}
We conduct experiments on three real-world datasets: Movielens1M \cite{maxwell2016movielens}, Amazon Electronics \cite{he2016ups} and Gowalla \cite{liang2016modeling}. Following \cite{he2020lightgcn,wang2019neural}, we perform a \textit{k}-core preprocessing step setting $k_{core} = 10$. We randomly split the interaction data of each user in train (80\%), validation (10\%) and test set (10\%), we use the validation data to determine the best algorithm hyperparameters, subsequently, we assess their final performance on the test set by training the models with both train and validation data. We compare the proposed algorithm with BPR \cite{rendle2012bpr} and LightGCN \cite{he2020lightgcn} as well as the spectral methods PureSVD \cite{cremonesi2010performance} and SGMC \cite{chen2021scalable} and the linear model EASE \cite{steck2019embarrassingly}. 
 The code used to produce the presented experiments is publicly available on github\footnote{https://github.com/damicoedoardo/PSGE}.

\subsection{Recommendation Performance}
To evaluate the algorithm's recommendation performance under the two different aspects of ranking and accuracy we report \textit{NDCG@20} and \textit{Recall@N} using two different cutoffs, $N=\{5,20\}$ and present the results in Table \ref{table:performance_table}.
Except for the NDCG on Movielens1M, where it is the second best performer, \psge{} gets the best results on the Recall and NDCG metrics for both cutoffs in all datasets studied, demonstrating its effectiveness in comparison to well-known, high-performing baselines from the graph convolution and spectral research domains. When compared to LightGCN, the model that inspired the study, \psge{} consistently outperforms it in all datasets, with a minimum gain of $11\%$ 
on the NDCG@20 on Movielens1M and a maximum increment of 
$29\%$ 
on NDCG@20 on Gowalla. We conclude that in the context of implicit interaction data, we can mimic the effect of graph convolution without resorting to a costly gradient-based optimisation approach. \psge{} corresponds to  the SGMC algorithm with the setting $\alpha = \beta = 0.5$. We can see that in all the datasets and for all metrics and cutoffs, the introduction of the two tuneable parameters accounting for the propensity scoring of users and items, is capable of delivering substantial improvements over its hypergraph counterpart formulation in which the exponent is set to  a fixed value. 
\subsection{Controlling Popularity Bias} \input{tex_figures/avg_popularity_exp} \psge{} reintroduces both user and item popularity to approximate the interaction matrix by rescaling the norm of the user and item embeddings by their respective degree (see Eq.
\ref{eq:psge_prediction}). We note that rescaling on the users has no influence on the ranking at prediction time, but rescaling on the items increases the popularity on the recommendations. This enables us to control the popularity in the predictions by trading it off against recommendation performance- we achieve this by changing the value of $\beta$ used to estimate $\hat{\mat{R}}$. 
To evaluate the algorithm's efficacy from this standpoint, in Fig. \ref{fig:pop-injection} we show the average popularity in the \psge{} prediction against the performance when the exponent $\beta$ (associated with the item degree rescaling) is varied. We also show a comparison to the behaviour of the baseline. The average popularity in the prediction is defined as the mean of the popularity of the items recommended, while the item popularity is defined as $pop_i = d_i/U$, where $d_i$ indicates the item degree. For clarity, we refer to $\tilde{\beta}$ as the manipulated parameter while $\beta$ refers to the value used in computing the normalised interaction matrix $\tilde{\mat{R}}$.
We vary $\tilde{\beta}$ in the range $[0, 1]$ with a step size of $0.1$. The mean popularity of the recommendation increases monotonically with $\Tilde{\beta}$, while the NDCG peaks at a value of $\tilde{\beta}$ close to $\beta$. On Movielens1M and Amazon the peak is observed exactly at $\tilde{\beta}$ = $\beta$, while on Gowalla we reach the best performance at $\tilde{\beta} = 0.3$ while $\beta = 0.4$. From the presented results we have empirically demonstrated how our algorithm can effectively trade off recommendation performance in favour of lowering popularity in the recommendations. It is worth mentioning that in all datasets, \psge{} recommendations associated with peak performance have lower average popularity when compared to the second best performing algorithm, highlighting how the algorithm is capable of generating high quality predictions. 

%% file: tex_tables/trainval_perf_table_2.tex
\begin{table*}[!t]
\caption{Recommendation performance. bold and underline indicate the first and the second best performing algorithms.}
\small
\setlength{\tabcolsep}{1.5pt}
\renewcommand{\arraystretch}{1.2}
\centering
\begin{tabular}{l@{\hspace{5pt}}ccc@{\hspace{5pt}}ccc@{\hspace{5pt}}ccc}

\multirow{3}{*}{\textbf{Model}} &

\multicolumn{3}{c}{\textbf{Ml1M}} &
\multicolumn{3}{c}{\textbf{Amazon}} &
\multicolumn{3}{c}{\textbf{Gowalla}} \\

\cmidrule(l{0pt}r{10pt}){2-4}
\cmidrule(l{0pt}r{10pt}){5-7}
\cmidrule(l{0pt}r{10pt}){8-10}
& 
 \multicolumn{1}{c}{NDCG} & 
 \multicolumn{2}{c}{Recall} &
 
 \multicolumn{1}{c}{NDCG} & 
 \multicolumn{2}{c}{Recall} &
 
 \multicolumn{1}{c}{NDCG} &
 \multicolumn{2}{c}{Recall}  \\ 
 
 & 
 @20 & @5 & @20 & 
 @20 & @5 & @20 & 
 @20 & @5 & @20 \\ \midrule
 
 BPR-MF &
 0.2602 & 0.1191 & 0.2756  &
 0.0439 & 0.0377 & 0.0928  & 
 0.1021 & 0.0763 & 0.1721  \\
 
  LightGCN &
 0.2679 & 0.1254 & 0.2898  &
 0.0446 & 0.0357 & 0.0956  & 
 0.1277 & 0.0980 & 0.2050  \\
 
 \midrule
 
     PureSVD &
 0.2621 & 0.1203 & 0.2755  & 
 0.0299 & 0.0229 & 0.0682  & 
 0.1162 & 0.0860 & 0.1836  \\
 
 EASE &
 \textbf{0.2969} & \underline{0.1415} & \underline{0.3164}  & 
 0.0509 & 0.0435 & 0.1028  & 
 0.1469 & 0.1114 & 0.2319 \\
 
  SGMC &
 0.2830 & 0.1369 & 0.3070  & 
 \underline{0.0528} & \underline{0.0443} & \textbf{0.1087}  & 
 \underline{0.1514} & \underline{0.1167} & \underline{0.2328}  \\
 
 \midrule

    
     
      
 
 
 
 

 PSGE &
 \underline{0.2951} & \textbf{0.1418} & \textbf{0.3230} &
 \textbf{0.0533} & \textbf{0.0458} & \textbf{0.1087} & 
\textbf{0.1641} & \textbf{0.1265} & \textbf{0.2519}\\
 
\midrule
\textbf{Statistics} & & & & & & & & & \\[-2pt]
\# users& \multicolumn{3}{c}{5 949}& \multicolumn{3}{c}{9 279}& \multicolumn{3}{c}{29 858}\\[-2pt]
\# items& \multicolumn{3}{c}{2 810}& \multicolumn{3}{c}{6 065}& \multicolumn{3}{c}{40 988}\\[-2pt]
\# inter& \multicolumn{3}{c}{571 531}& \multicolumn{3}{c}{158 979}& \multicolumn{3}{c}{1 027 464}\\




\midrule

\end{tabular}
\label{table:performance_table}
\end{table*}

%% file: tex_figures/avg_popularity_exp.tex
\begin{figure*}
    \centering
    \subfigure[Movielens1M]{\includegraphics[width=0.32\columnwidth]{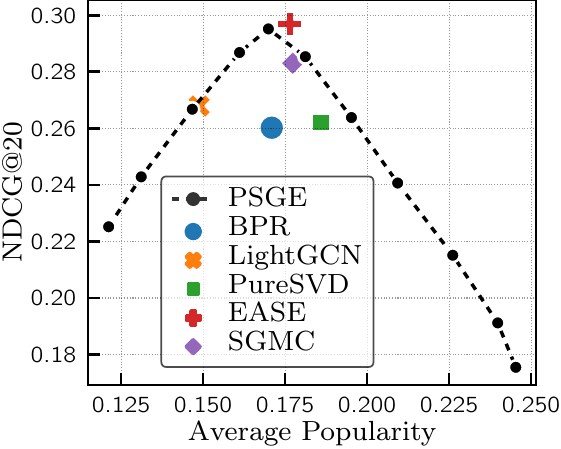}}
    \subfigure[Amazon]{\includegraphics[width=0.32\columnwidth]{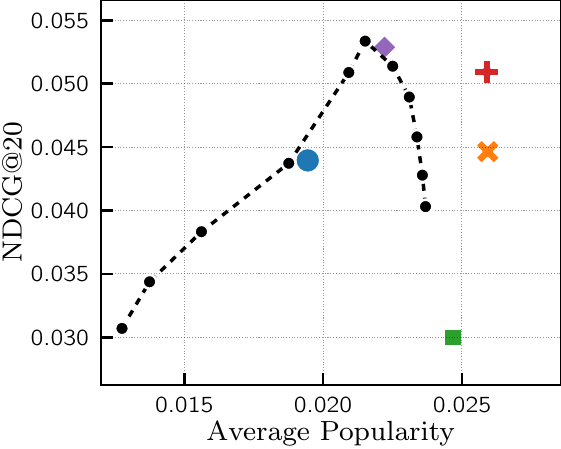}}
    \subfigure[Gowalla]{\includegraphics[width=0.32\columnwidth]{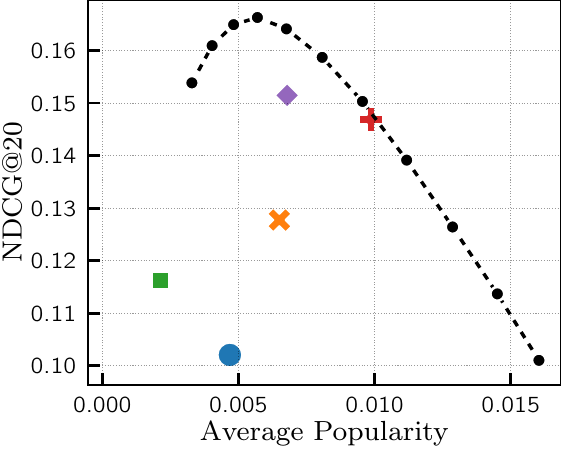}}
    \caption{Manipulation of the hyperparameter $\beta$ regulating the item degree norm rescaling. 
    The tradeoff between accuracy and popularity in the predictions is reported plotting the NDCG@20 against the average popularity on the recommendations.}
    \label{fig:pop-injection}
\end{figure*}

%% file: 05-Conclusion.tex
\section{Conclusion}
We presented a study on the graph convolution approach employed by LightGCN proving how the convolution acts as a fixed, high-pass filter in the spectral domain induced by the normalised adjacency matrix. We presented a detailed explanation of why this operation is beneficial to the top-$N$ recommendation problem. 
Exploiting this spectral interpretation, we presented a scalable spectral algorithm based on the singular value decomposition of the propensity weighted interaction matrix. We empiracally showed how the presented model is able to emulate the behaviour of the light convolution by achieving better performance than LightGCN, requiring only a fraction of the training time and enabling the control of the tradeoff between accuracy and popularity on the set of provided recommendations.